# Towards optimal single-photon sources from polarized microcavities

Hui Wang[1,2,*], Yu-Ming He[1,2,*], Tung Hsun Chung[1,2], Hai Hu[3], Ying Yu[4], Si Chen[1,2], Xing Ding[1,2], Ming-Cheng Chen[1,2], Jian Qin[1,2], Xiaoxia Yang[3], Run-Ze Liu[1,2], Zhao-Chen Duan[1,2], Jin-Peng Li[1,2], Stefan Gerhardt[5], Karol Winkler[5], J. Jurkat[5], Lin-Jun Wang[1], Niels Gregersen[6], Yong-Heng Huo[1,2], Qing Dai[3], Siyuan Yu[4], Sven Höfling[5,7], Chao-Yang Lu[1,2,*], Jian-Wei Pan[1,2,*]

cylu@ustc.edu.cn (CYL), pan@ustc.edu.cn (JWP)

[1] Shanghai branch, National Laboratory for Physical Sciences at Microscale and Department of Modern Physics, University of Science and Technology of China, Shanghai 201315, China
[2] CAS Centre for Excellence and Synergetic Innovation Centre in Quantum Information and Quantum Physics, University of Science and Technology of China, Hefei, Anhui 230026, China
[3] Nanophotonics Research Division, CAS Center for Excellence in Nanoscience, National Center for Nanoscience and Technology, Beijing, 100190, China
[4] State Key Laboratory of Optoelectronic Materials and Technologies, School of Electronics and Information Technology, Sun Yat-sen University, Guangzhou 510275, China
[5] Technische Physik, Physikalisches Instität and Wilhelm Conrad Röntgen-Center for Complex Material Systems, Universitat Würzburg, Am Hubland, D-97074 Würzburg, Germany
[6] DTU Fotonik, Department of Photonics Engineering, Technical University of Denmark, Ørsteds Plads 343, DK-2800 Kongens Lyngby, Denmark
[7] SUPA, School of Physics and Astronomy, University of St. Andrews, St. Andrews KY16 9SS, United Kingdom

**Abstract:**

**An optimal single-photon source should deterministically deliver one and only one photon at a time, with no trade-off between the source's efficiency and the photon indistinguishability. However, all reported solid-state sources of indistinguishable single photons had to rely on polarization filtering which reduced the efficiency by 50%, which fundamentally limited the scaling of photonic quantum technologies. Here, we overcome this final long-standing challenge by coherently driving quantum dots deterministically coupled to polarization-selective Purcell microcavities—two examples are narrowband, elliptical micropillars and broadband, elliptical Bragg gratings. A polarization-orthogonal excitation-collection scheme is designed to minimize the polarization-filtering loss under resonant excitation. We demonstrate a polarized single-photon efficiency of 0.60(2), a single-photon purity of 0.991(3),**

**and an indistinguishability of 0.973(5). Our work provides promising solutions for truly optimal single-photon sources combining near-unity indistinguishability and near-unity system efficiency simultaneously.**

Single photons are appealing candidates for quantum communications[1,2], quantum-enhanced metrology[3,4] and quantum computing[5,6]. In view of the quantum information applications, the single photons are required to be controllably prepared with a high efficiency into a given quantum state. Specifically, the single photons should possess the same polarization, spatial mode, and transform-limited spectro-temporal profile for a high-visibility Hong-Ou-Mandel-type quantum interference[1,7].

Self-assembled quantum dots show so far the highest quantum efficiency among solid-state quantum emitters and thus can potentially serve as an ideal single-photon source[8-15]. There has been encouraging progress in recent years in developing high-performance single-photon sources[11]. Pulsed resonant excitation on single quantum dots has been developed to eliminate dephasing and time jitter, which created single photons with near-unity indistinguishability[15]. Further, by combining the resonant excitation with Purcell-enhanced micropillar[16,17] or photonic crystals[18,19], the generated transform-limited[20,21] single photons have been efficiently extracted out of the bulk and funneled into a single mode at far field.

Despite the recent progress[16-22], the experimentally achieved polarized-single-photon efficiency (defined as the number of single-polarized photons extracted from the bulk semiconductor and collected by the first lens per pumping pulse) is ~33% in ref. 16 and ~15% in ref. 17, still fell short of the minimally required efficiency of 50% for boson sampling to show computational advantage to classical algorithms[23], and was below the efficiency threshold of 67% for photon-loss-tolerant encoding in cluster-state models of optical quantum computing[24]. It should be noted that a <50% single-photon efficiency would render nearly all linear optical quantum computing schemes[1,5] not scalable.

The indistinguishable single-photon source efficiency remained a persistent problem for years, due to two main reasons. First, in the previous experiments[16,20,22] that relied on the optically bright, doubly degenerate transitions in singly-charged quantum dots, the optical selection rule dictates that the resonance fluorescence photons are randomly right or left circularly polarized. For many quantum information applications which requires input single photons in a fixed polarization state, passive polarization filtering is needed to project into a single polarization, which reduces the system efficiency by at least a factor of 2.

The other reason is associated with the use of the resonant excitation method that requires suppression of the pump laser light having the same frequency as the single photons. So far, the most effective method is polarization filtering[15,25], where a linearly polarized laser excites the quantum dot, and a cross polarizer in the collection arm extinguishes the scattering laser background. However, such a polarization filtering again reduces the system efficiency of the quantum-dot single photons by at least a factor of 2. One possible remedy is to excite the quantum dot from the side into the waveguide mode of the micropillar and collect single photons from the top. Yet, the spatially-orthogonal excitation-collection methods[26,27] were only attempted in low-Purcell and non-polarized quantum dot devices and didn't realize background-free, high-performance single-photon sources. Therefore, overcoming the outstanding challenge of 50% efficiency loss has remained the most difficult and final challenge for an ultimate high-performance single-photon source, both theoretically and technologically.

**Theoretical scheme**

To develop a polarized single-photon source with simultaneously near-unity system efficiency and near-unity indistinguishability, we (1) break the original polarization symmetry of the quantum dot emission, and (2) develop a new way for background-free resonance fluorescence without sacrificing the system efficiency. Here, for the first time, we put forward a feasible proposal that kills the two birds with one stone and report its experimental demonstrations. A general framework of our protocol is shown in Fig. 1a. The key idea is to couple a single quantum dot to a geometrically birefringent cavity in the Purcell regime. The asymmetric microcavity is designed such that it lifts the polarization degeneracy of the fundamental mode and splits it into two orthogonal linearly polarized—horizontal (*H*) and vertical (*V*)—modes, with a cavity linewidth of $\delta\omega$ and a frequency separation of $\Delta\omega$. Suppose a single-electron charged quantum dot, which is a degenerate two-level system, is brought into resonance with the cavity *H* mode, and off-resonant with the cavity *V* mode with a detuning of $\Delta\omega$. The cavity redistributes the spontaneous emission rate of the quantum emitter into the *H* and *V* polarizations with a ratio of $1+4(\Delta\omega/\delta\omega)^2:1$ (see Supplementary information). For a series of realistic Purcell factors, the extraction efficiency of polarized single photons is plotted in Fig. 1b. For example, with a Purcell factor of 20 and $\Delta\omega/\delta\omega=3$, the polarized single-photon extraction efficiency is 93%.

Having singled out a polarized two-level system, we further need to develop a way for resonantly driving the quantum dot transition and for obtaining near-background-free collection of the resonance fluorescence, which is another outstanding challenge by its own. Our method is compatible with the cross-polarization technique but has little loss of single-photon system efficiency. The excitation laser's polarization is set to be *V*, while in the output, an *H* polarizer—aligned with the dominant polarization of the Purcell-enhanced single photons—is used to extinguish the pump laser scattering. Note that our protocol favors a suitable $\Delta\omega/\delta\omega$ ratio, practically in the range of 1.5-3. As the *V*-polarized excitation laser off-resonantly couples to the *V*-cavity mode at the emitter's transition (Fig. 1a), the π-pulse driving power needs to be moderately stronger compared to the case with an isotropic microcavity (the increasing factor is calculated and plotted in Fig. 1b). For example, at $\Delta\omega/\delta\omega=2.5$, the excitation laser power is estimated to be ~7 times higher.

The protocol is applicable in many photonic structures[8-11] such as micropillars[16,17,20], bullseyes[28,29], microdisks[30], nanowires[31,32], and photonic crystals[18,19]. Here, we demonstrate our protocol by resonant excitation of single quantum dots deterministically embedded in two types of polarized microcavities—elliptical micropillars (narrowband) and elliptical Bragg gratings (broadband)—for the generations of single-polarized single-photon sources with both high system efficiency and near-unity indistinguishability.

**Elliptical micropillar**

The GaAs/InAs micropillars with elliptical cross section were first studied by Gayral *et al.* in 1998, who observed a splitting of the degenerate fundamental modes of the cavity[33]. It was shown that the two split modes were linearly polarized, aligned in parallel with the major and minor axis of the elliptical cross section[34,35]. Single GaAs/InAs quantum dots embedded in such elliptical micropillars showed polarization-dependent Purcell enhancement[36,37] and single photons were preferentially generated in a single polarized state[33-39] with up to 44% extraction efficiency[39]. However, all previous work with elliptical micropillars was performed with non-resonant excitation, which degraded the purity of the emitted photons. Moreover, neither high collection efficiency nor high indistinguishability was realized in the first generation of elliptical micropillar devices.

In this work, we use two-color photoluminescence imaging to determine the position of single quantum dots preselected with bright emission and narrow linewidth[29]. With suitably low quantum dot density (~$4\times10^7$ cm$^{-2}$), the wide-field optical imaging method can obtain a position accuracy of ~22 nm (see Fig. 2a and Supplementary Information). This allows us to deterministically etch the micropillar with the quantum dot in the center of the cavity, which is important for an optimal emitter-cavity coupling.

The micropillar devices in our experiment have elliptical cross sections (see Fig. 2a) with major (minor) axis diameter of 2.1 μm (1.4 μm). The elliptical micropillar is characterized using non-resonant excitation with a ~780 nm c.w. laser at high power. Fig. 2b shows the two nondegenerate fundamental cavity modes[33,35-39], at 896.54 nm (labelled as $M_1$) and 897.04 nm (labelled as $M_2$), respectively, with a splitting of 183 GHz. The $M_1$ and $M_2$ modes correspond to the minor and major axis, with quality factors of 4075 and 5016, respectively. A modest reduction of the quality factor of $M_1$ compared to $M_2$ is due to the smaller micropillar diameter. Polarization-resolved measurements (see Fig. 2c) confirm that the polarization of $M_1$ ($M_2$) is parallel to the minor (major) axis which we label as *V* (*H*), with a high degree of polarization of 99.7% (99.6%), which confirms symmetry-broken, highly-polarized nature of the microcavity.

**Elliptical Bragg grating**

The circular Bragg grating bullseye microcavities, which tightly confine the light in a sub-λ transverse plane, were previously fabricated on quantum dots[28,29] and diamond nitrogen-vacancy centers[40] for enhancing their collection efficiencies. Recently, a thin low-refractive-index $SiO_2$ layer and a gold mirror[31,41-47] was added at the bottom of the bullseye to redirect the downward propagating light towards the top, thus improving the collection efficiency to near unity. Here, for the first time, we break the polarization symmetry of the circular Bragg grating and design it as a geometrically birefringent Purcell cavity.

The spectral resonance of the bullseye cavity is strongly sensitive to the radius of the central disk and the grating period. It was previously shown that a 1-nm change of the central disk radius (grating period) causes a 1.14 nm (0.25 nm) shift of the cavity mode[46]. Thus, a 1% ellipticity, which is 23 times smaller than that of our elliptical micropillar, is sufficient to induce a suitable cavity mode splitting. This is favorable for producing a near-Gaussian far-field emission profile.

Assisted by finite-different time-domain simulations, we design the elliptical Bragg grating (see Fig. 2d) featuring a central elliptical disk with a major (minor) axis of 770 nm (755 nm), surrounded by an elliptical grating with a period of 380 nm (372 nm) for the major (minor) axis, with fully-etched 100 nm-width trenches (see Supplementary information for detailed fabrication process). The cavity mode of our fabricated bullseye device (see Fig. 2e) splits into a doublet (labelled as $B_1$ and $B_2$, in *V* and *H* polarization, respectively). The splitting is 2.8 THz, which is 1.5 (1.3) times larger than the linewidth of the $B_1$ ($B_2$). The broadband nature of the bullseye geometry greatly facilitates the emitter-cavity spectral alignment. As shown in Supplementary Fig. 7, the design can ideally yield a high Purcell factor of >20 in a few nm wavelength range.

**Polarized indistinguishable single photons**

The samples are placed inside a bath cryostat with the lowest temperature of 1.5 K. A confocal microscope is used to excite the quantum dot and collect the emitted single photons. Following our protocol shown in Fig. 1a, driven by a *V*-polarized laser, the *H*- polarized single photons from the Purcell-enhanced $M_2$ cavity spectrally resonant with a charged quantum dot are created and collected into the output of the microscope with a ~$10^7$:1 cross-polarization extinction of the scattered laser background. We first study the emitter-cavity coupling and measure the Purcell factors in both devices. With time-resolved resonance fluorescence measurements (see Fig. 2f), the radiative lifetime for the single quantum dots coupled to the elliptical micropillar and the elliptical Bragg grating are ~61.0(1) ps and ~69.1(1) ps, respectively, which is ~17.8 and ~15.7 times shorter than the average lifetime (~1.09 μs) of more than 20 quantum dots in the slab from the same area. Such high Purcell factors serve to efficiently funnel the spontaneous emission into a single output mode and additionally to reduce the influence of dephasing on the indistinguishability[48].

Figure 3a shows the detected resonance fluorescence single photons as a function of the driving field amplitude, and a full Rabi oscillation curve is observed. With a pumping repetition rate of 76 MHz and using a π pulse, about 13.7 million photon counts per second are detected by a single-mode fiber-coupled superconducting nanowire single-photon detector with an efficiency of ~76%. A comparison experiment is performed where we excite the dot with a *H*-polarized laser and collect *V*-polarized single photons (Fig. 1a). As shown in Fig. 3b, using a π pulse, only about 0.54 million photons per second are detected. We estimate a degree of polarization—defined as $(I_H - I_V)/(I_H + I_V)$, where $I_H$ ($I_V$) denotes the detected intensity of *H*- (*V*-)polarized photons—of 92.3% for the generated single photons from the polarized micropillar device. Thus, the single-photon source suffers a loss of 3.8% due to polarization only, whereas the previous resonance fluorescence experiments lost at least 50% photons in polarization filtering[16,17,20,22]. Similarly, in the elliptical bullseye device, we detect 12.4 million polarized resonance fluorescence for a π pulse. To our knowledge, this is the first time that resonance fluorescence is observed for the quantum dot two-level systems in the bullseye membrane structures.

A detailed photon loss budget analysis for both devices, including quantum radiative efficiency, blinking, collection optics, optical path transmission efficiency, and single-mode fiber coupling, is presented in the Supplementary Information. For the elliptical micropillar-quantum dot device, the polarized single-photon efficiency is measured to be 0.60(2), where the dominant loss mechanisms are imperfect sidewall scattering[49] (~22.8%), mode leakage (~5.3%), and imperfect internal quantum efficiency (including the excited-state preparation efficiency at the π pulse and the radiative efficiency of the quantum dot) which is estimated to be ~82%. For the elliptical bullseye devices, the polarized single-photon efficiency is 0.56(2), where the loss is mainly due to the quantum dot blinking (35%), as the dot is close to the etched surface with a distance of 62.5 nm. The blinking can be reduced in the future by surface passivation and by applying an electric field.

The purity of the elliptical micropillar single-photon source is characterized with a Hanbury Brown and Twiss setup. As displayed in Fig. 4a, the measured second-order correlation data shows $g^2(0) = 0.025(5)$ at the zero-time delay. The imperfection of the measured $g^2(0)$ is mainly due to laser leakage. The photon indistinguishability is measured using a Hong-Ou-Mandel interferometer with the time separation between the two consecutive excitation laser pulses—and thus the two emitted single photons—set at 13 ns. Figure 4b shows the photon correlation histograms of normalized two-photon counts for orthogonal and parallel polarizations. The observed contrast of the counts for the two cases at zero delay can be used to extract a raw two-photon interference visibility of 0.913(5). Taking into account of the imperfect single-photon purity and the unbalanced (47:53) beam splitting ratio in the optical setup, we calculate a corrected photon indistinguishability of 0.975(6). For the elliptical bullseye device, similar results are observed: after send the photons through a 5-GHz etalon to filter out the phonon sideband (which meanwhile reduces the single-photon count by ~30% due to the filtering and etalon transmission rate), a single-photon purity of 0.991(3) and an indistinguishability of 0.951(5) for two single photons separated by 13 ns is presented in the Supplementary Information. These results show that high efficiency, purity and indistinguishability can be simultaneously combined with high degrees of polarization in a coherently driven single-photon device.

**Discussion and conclusion**

In summary, we have theoretically proposed and experimentally demonstrated near-optimal single-photon sources operating in a single polarization combined with simultaneous high purity, indistinguishability and efficiency by resonantly driving single quantum dots coupled to polarized microcavities, thus overcoming the long-standing problem of 50% photon loss. Our design is compatible with the mature cross-polarization set-up and does not require complicated sample fabrications, and side excitation set-up etc. This has enabled the creation of the brightest polarized sources of single indistinguishable photons to date in all physical systems. The generality and versatility of our protocol have been demonstrated in both narrowband and broadband cavities, which, interestingly, have their own pros and cons. The bullseye cavity is robust to fabricate, requires negligible ellipticity and features tighter confinement. The micropillar cavity requires a more precise matching of the microcavity-emitter spectral resonance than the bullseye device, but also comes with a benefit that the phonon emission can be suppressed due to its narrowband nature. It will be interesting in the future to combine these two nanostructures together[50] to enjoy both their merits including tight confinement, minimal scattering loss, and suppression of phonon-assisted emission. Further optimizations on the sample fabrication, electrical gating, surface passivation[51] and single-mode fiber coupling are expected to deliver near-optimal single photons[52], the central resources for optical quantum information technologies[53,54].

**Figure captions:**

**Figure 1: The theoretical scheme of a polarized single-photon source by resonantly pumping a quantum emitter in a birefringent microcavity**. **a.** The asymmetric cavity supports two-fold non-degenerate cavity modes, one in horizontal ($H$) polarization and the other one in vertical ($V$) polarization. We bring the emitter into resonance to the $H$ mode, at which single photons are preferentially prepared due to polarization-selective Purcell enhancement. The $V$ cavity mode, suitably detuned from the emitter, serves for laser excitation. A $V$ polarized laser pulse is weakly coupled to the $H$ mode due to a small overlap between the two cavity modes, thus can pump the two-level system to its excited state with a higher power. **b.** The system efficiency of preparing single polarized single photons as a function of the ratio of the cavity splitting to the cavity linewidth. Examples with four different Purcell factors are plotted. The dashed line shows the increased factor of the pump laser power required for a deterministic $\pi$ pulse, compared to the case of micropillar with circular cross section.

**Figure 2: Characterization of the elliptical micropillar and elliptical Bragg grating**. **a**. An illustration of InGaAs quantum dot-elliptical micropillar device used in this work, which has a major (minor) axis of 2.1 μm (1.4 μm). The quantum dot is sandwiched

between 25.5 (15) λ/4-thick AlAs/GaAs mirror pairs forming the lower (upper) distributed Bragg reflectors. **b**. Two fundamental modes of the elliptical micropillar, $M_1$ and $M_2$, with a splitting of 171 GHz. The linewidths of the $M_1$ and $M_2$ are 74 GHz and 62 GHz, respectively. The quantum dot is resonant with the $M_2$ at a temperature of 4 K. **c.** Polarization-resolved measurement of the two cavity modes, which are perpendicular with each other. The degree of polarization of the $M_1$ and $M_2$ modes is 99.7% and 99.6%, respectively. **d**. A schematic structure of the elliptical Bragg grating, which consists of a central elliptical disk, surrounding elliptical grating, and fully-etched trenches. A thin low-refractive-index $SiO_2$ layer and a gold mirror were added at the bottom of the Bragg grating, followed by a 500-nm thick SU-8 and a 500-μm silicon substrate. The red dot indicates the position of the coupled quantum dot. **e.** Two non-degenerated modes of broadband bullseye cavity, $B_1$ and $B_2$, with a splitting of 2.8 THz, which are 1.5 (1.3) times larger than the linewidth of the $B_1$ ($B_2$). The investigated quantum dot is resonant with the $B_2$. **f.** Radiative lifetime of the quantum dots coupled to the bullseye cavity (red) and in micropillar cavity (blue), which is 69.1 (1) ps and 61.0 (1) ps, respectively, measured using a superconducting nanowire single-photon detector with a time resolution of 20 ps. The inset is the data of the lifetime (~1.1 ns) of the quantum dots in slab from the same area.

**Figure 3: The deterministic generation of polarized single photons under resonant excitation. a**, The polarization of the excitation laser is set to be parallel to the minor axis, and the output polarizer is aligned parallel to the major axis, that is, orthogonal to the input laser polarization in order to extinguish the laser background. The measured pulsed resonance fluorescence single photon counts are plotted as a function of the laser power, which shows a clear Rabi oscillation. Under π pulse at ~63 nW, we observe ~13.7 million single-photon counts per second by a superconducting nanowire single-photon detector. **b,** For a controlled experiment, we exchange the polarizations of the excitation laser and the collected single photons. In this case, the single photons are suppressed by the cavity. Under π pulse at ~5 nW, only 0.54 million single photons per second are detected.

**Figure 4: The single-photon purity and indistinguishability. a.** Measurement of the second-order correlation function, which gives $g^2(0) = 0.025(5)$. **b**, Characterization of photon indistinguishability in a Hong-Ou-Mandel interferometer with a time delay of 13 ns. A significant count suppression at zero delay is observed when the two photons are in parallel polarization (red dots) compared to the case when the two photons are in cross polarization (blue circles). After correction, the calculated indistinguishability is 0.975(6).

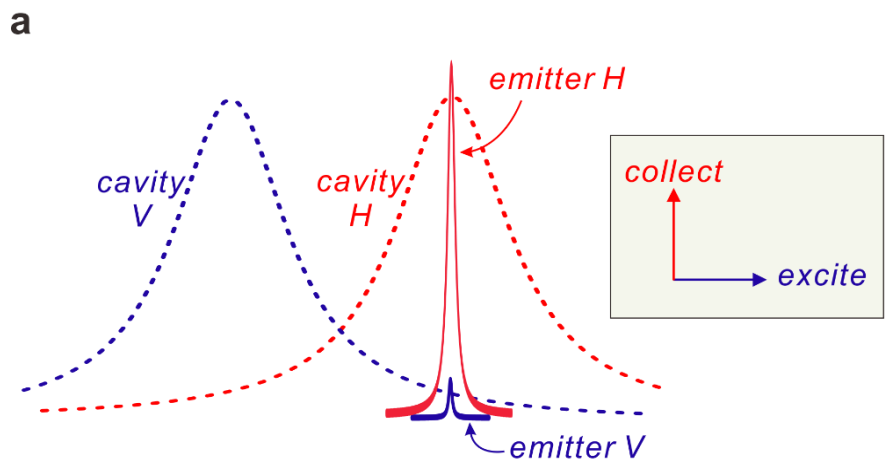 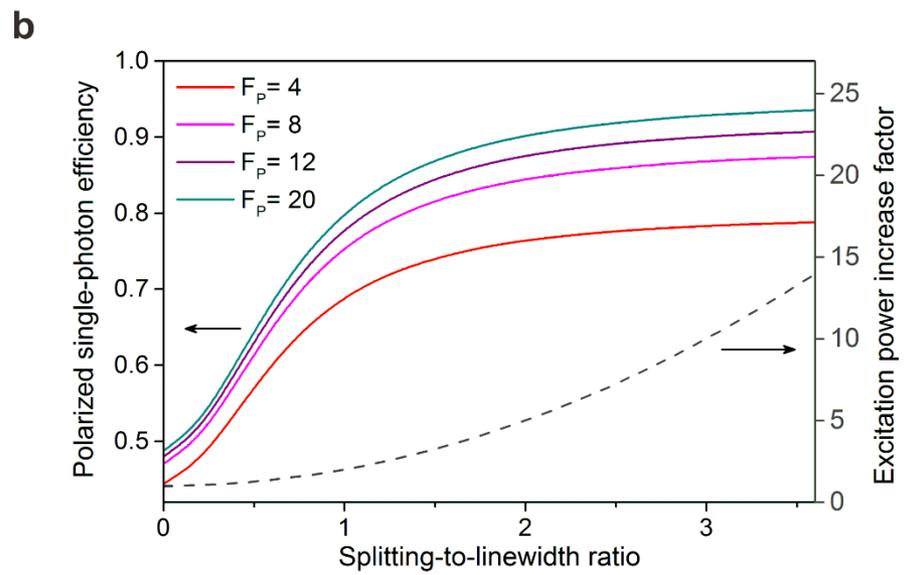

Figure 1

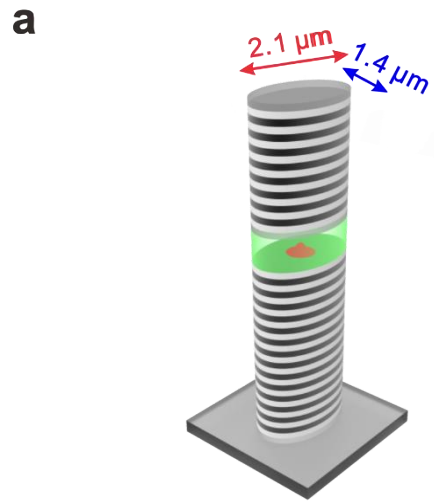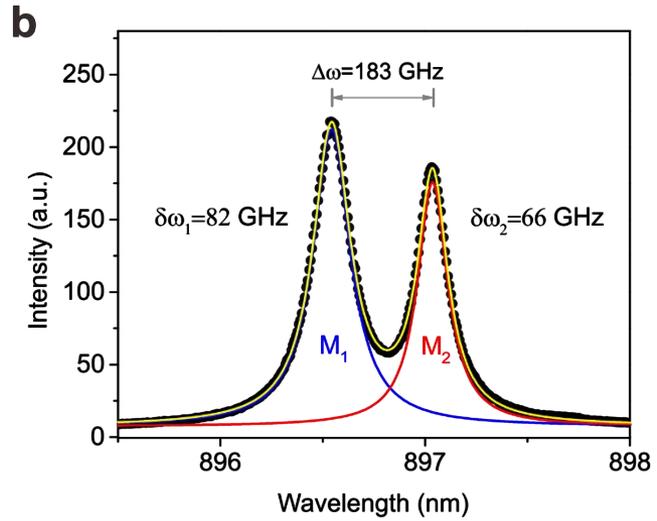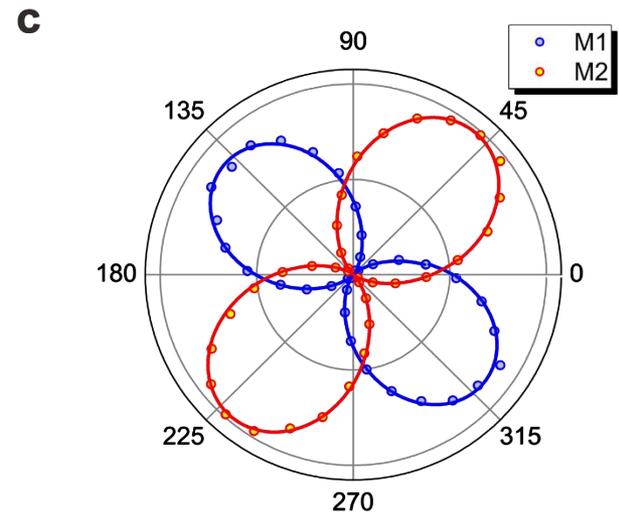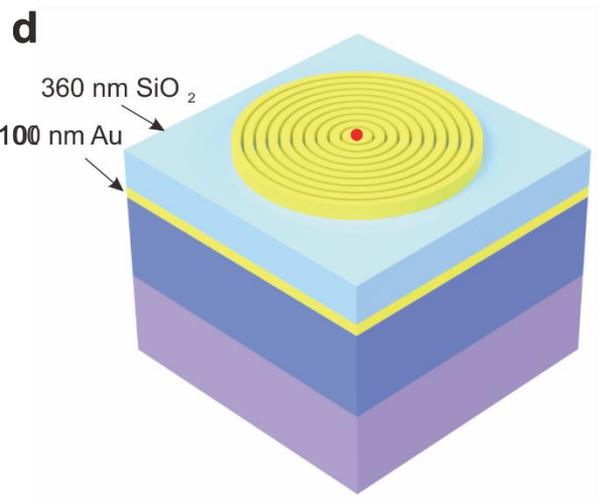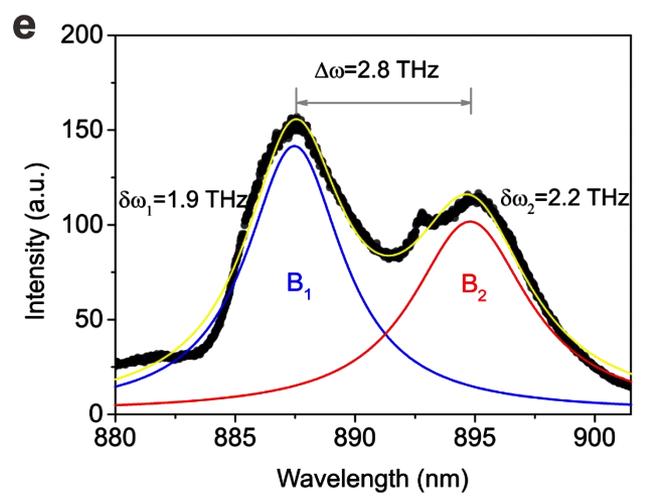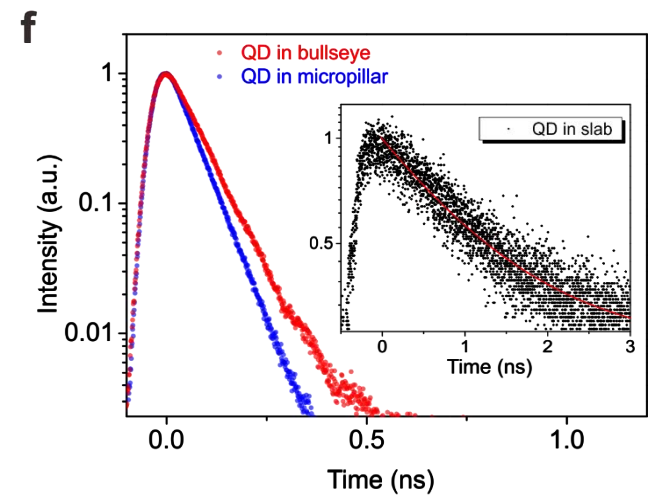

Figure 2

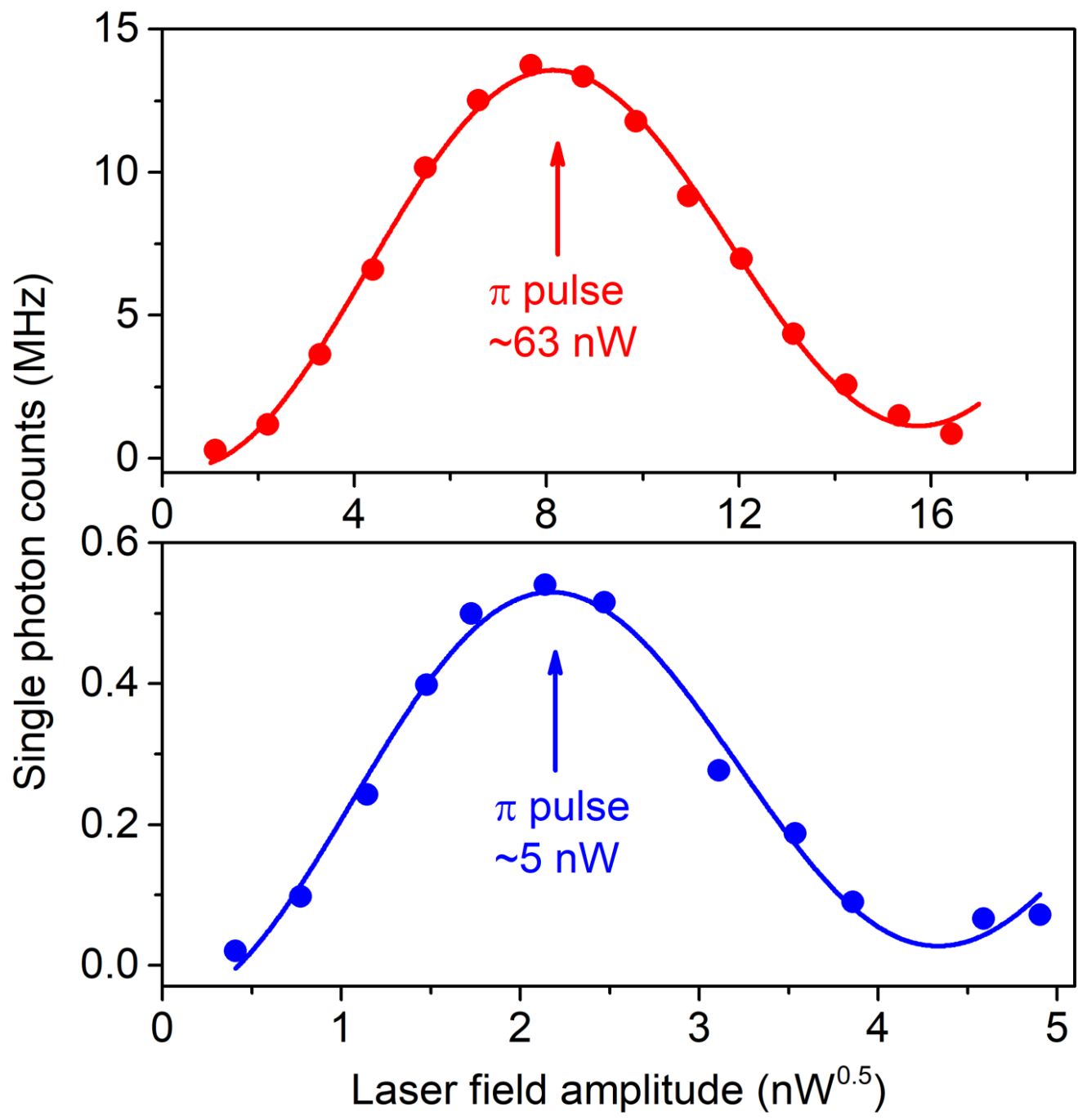

Figure 3

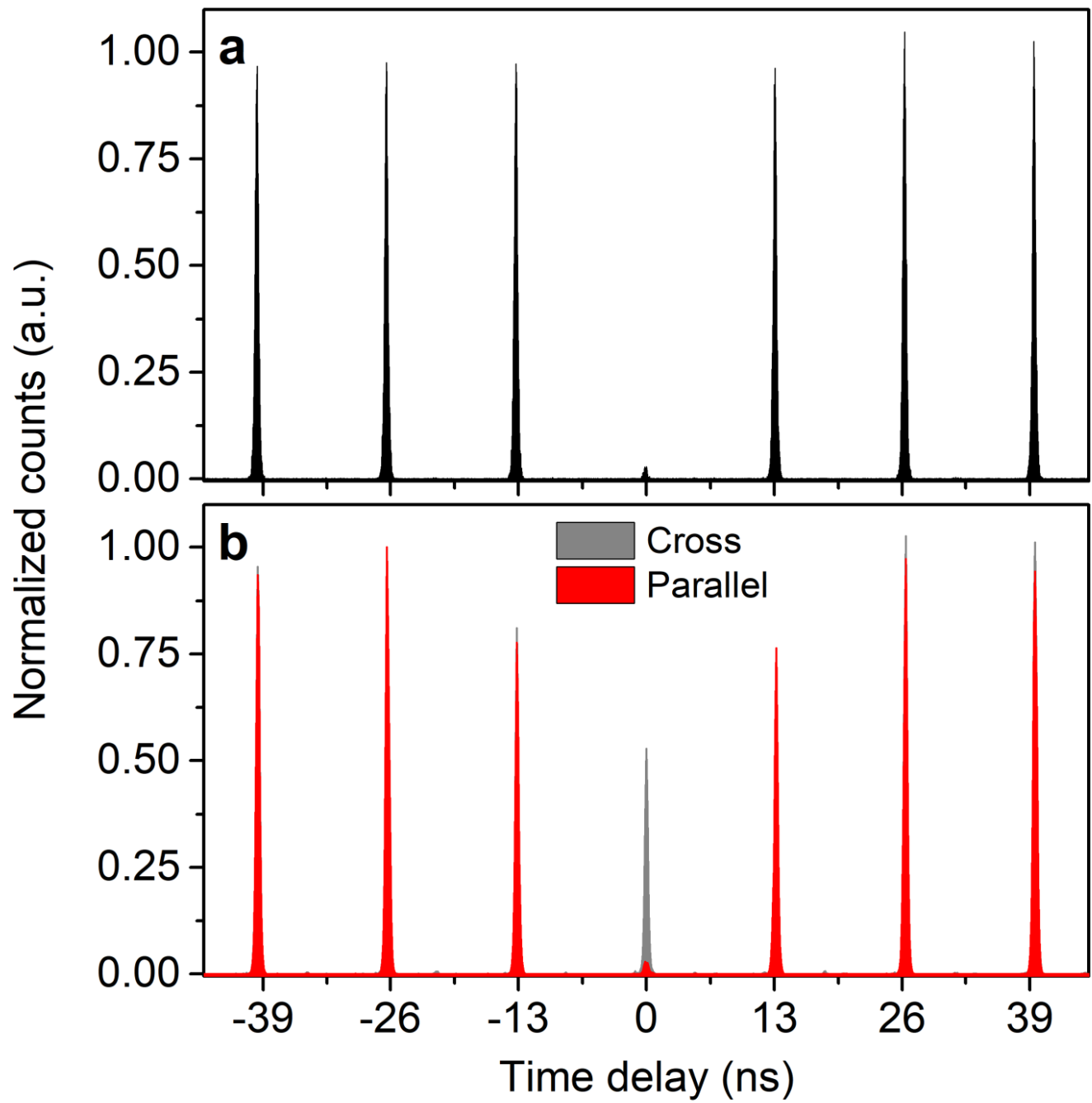

Figure 4

# Supplementary information

## A blueprint for all-perfect single-photon devices

In the diagram shown in Supplementary Fig. 12, we put forward a realistic blueprint (or recipe) towards engineering of all-perfect single-photon devices. This diagram involves a series of key steps; all these steps must be successfully implemented to deliver final working devices. Special attention should be paid which are listed below.

1. It started from molecular beam epitaxy growth of self-assembled quantum dots. The quantum dots should have near-unity quantum efficiency in the single-photon emission. This means that atomic defects need to be minimized during the growth.
2. For the fabrication of micropillars cavity later on, suitable layers of distributed Bragg reflector (DBR) mirrors should be monolithically grown below and above the quantum dot layer. The numerical simulation[11] has suggested that a cavity Q value of ~10,000 would be a sweet point to obtain both high efficiency and indistinguishability.
3. For the optical imaging and deterministic positioning of the quantum dots, the as-grown sample should have a low density so that the quantum dots are sufficiently separated.
4. Appropriate wide-field imaging (or other methods) should be performed and ensure a <20 nm positioning precision such that the dots are spatially maximally coupled to the cavities.
5. Special attention needs to be paid to the shift of both the DBR cavity resonance frequency and the quantum dot emission wavelength before and after the micropillar etching. Such shifts need be taken into account in the previous steps when pre-selecting good quantum dot candidates with bright and narrow line emissions.
6. Predict the final micropillar Q value. Based on the Q, empirically determine an appropriate major and minor axis diameter and ellipticity for polarization-controlled single photon emission.
7. Pay special attention to micropillar etching to ensure a smooth sidewall to minimize the scattering loss.
8. To stabilize the electric field fluctuation around the quantum dot and to tune the wavelength of the quantum dot emission, applications of surface passivation[12] and electric gate have been proven very helpful[13].

9. Resonantly and efficiently drive the quantum dot two-level system using an ultrafast laser pulse. The pulse duration should be much shorter than the radiative decay time of the two-level system to minimize the time jitter of the emitted single photons[14].

10. Achieve a sufficient suppression of the laser leakage using a combination of techniques including cross-polarization (only by using the method reported in our current paper is the single-photon efficiency not sacrificed), spatially orthogonal excitation (from side) and collection (from top), and/or spectrally orthogonal two-color resonant excitation[15].

**Supplementary References:**

**Supplementary Figure Captions:**

**Supplementary Fig. 1: (a)** Schematic of the photoluminescence imaging setup used for determining QD locations for deterministic elliptical micropillar-QD device fabrication. Illumination of the alignment marks uses a 780 nm LED. The unwanted light entering the EMCCD camera is removed by a long-pass filter (LPF) that blocks wavelengths below 715 nm and a notch filter (NF) that blocks wavelengths between 810 and 880 nm.

**Supplementary Fig. 2: (a)** Image acquired when only a 635-nm red LED is used for excitation of the QDs. **(b)** Horizontal line cut through the image from (b), along the dashed red line, showing the profile of the QD emission (black dots) and its Gaussian fit (red line) with an uncertainty of 2.5 nm. **(c)** Acquired image when only a 780-nm LED is used to illuminate the metallic alignment marks. **(d)** Horizontal line cut through the image from (d), along the dashed red line. **(e)** Image of the elliptical micropillars after subsequent fabrication using aligned e-beam lithography.

**Supplementary Fig. 3: (a)** A typical quantum dot imaged by EMCCD. **(b)** A 2D Gaussian fitting of the image in the **a**. **(c)** A typical image of the alignment mark to determine the position of the cross. **(d)** A 1D Gaussian fitting of the alignment mark.

**Supplementary Fig. 4:** Far-field intensity distribution of the elliptical micropillar cavity.

**Supplementary Fig. 5:** Top view of a scanning electron microscope image of the EBG cavity in X-Y plane.

**Supplementary Fig. 6: (a)** Polarization-resolved measurement of the two fundamental modes. The investigated QD is resonant with the cavity mode $B_2$. **(b)** Measured cavity modes of EBG cavities with different minor/major value from 0.92 to 1. The blue (red) dot corresponds to the cavity mode whose polarization is parallel with the major (minor) axis.

**Supplementary Fig. 7: (a)** Simulated intensity far-field distribution with a dipole along the minor axis. **(b)** Simulated extraction efficiency using an objective lens with NA=0.65, and the simulated Purcell factor of an EBG cavity with an ellipticity of 0.01.

**Supplementary Fig. 8: (a)** Simulated Purcell factors with different $b/a$ value from 0.94 to 1. **(b)** The peak position of the major and minor modes with different $b/a$. **(c)** Simulated Purcell factor with different thickness of $SiO_2$.

**Supplementary Fig. 9:** *H*-polarized single-photon counts under resonant excitation of a *V*-polarized pulse laser. At π pulse of 1.7 nW, we directly detected 12.4 M pure *H*-polarized single photons (red dots) using a superconducting nanowire single-photon detector.

**Supplementary Fig. 10: (a)** The single-photon purity is derived from the second-order correlation function. Our experiment shows a $g^2(0)=0.009(1)$, corresponding to a single-photon purity of above 99%. **(b)** The photon indistinguishability is characterized by a Hong-Ou-Mandel interferometer with a time delay of 13 ns. The peak of co-polarization (red) at zero delay is strongly suppressed with respect to the one with cross polarization (gray). After correction, the calculated indistinguishability is 95.1%. **(c)** Hong-Ou-Mandel interference of two successively emitted photons with a time delay of 2 ns. The extracted indistinguishability is 97.3%.

**Supplementary Fig. 11:** Blinking of the investigated QD in EBG cavity.

**Supplementary Fig. 12:** A blueprint for all-perfect single-photon devices.

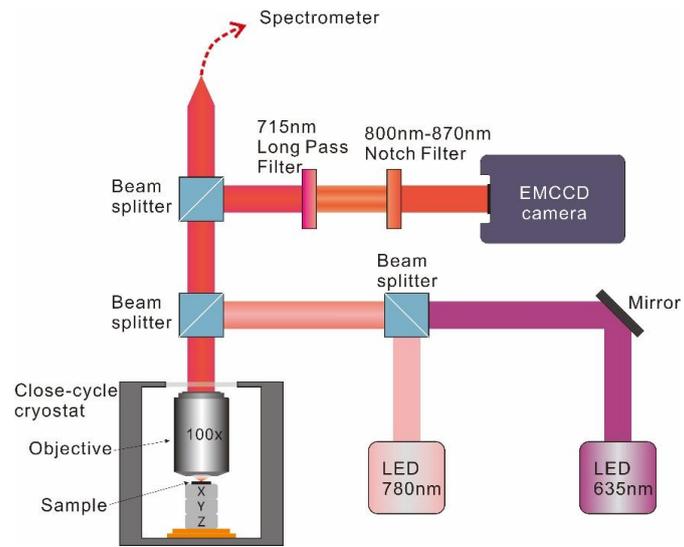

Supplementary Fig. 1

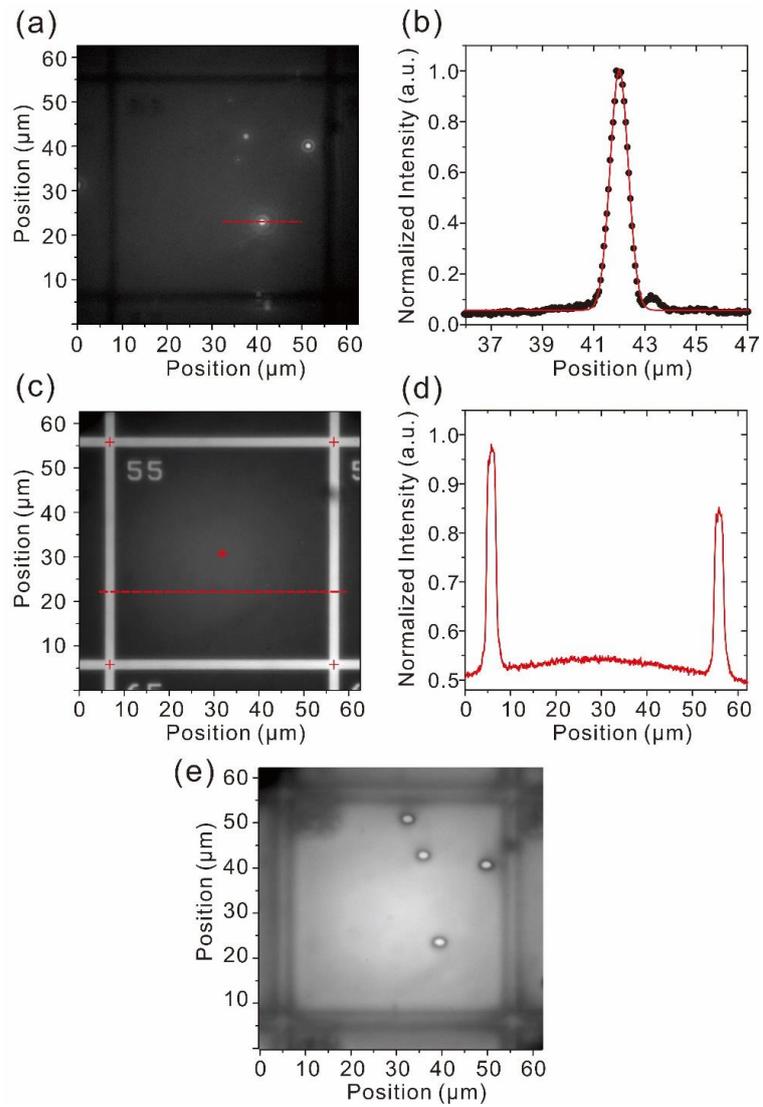

Supplementary Fig. 2

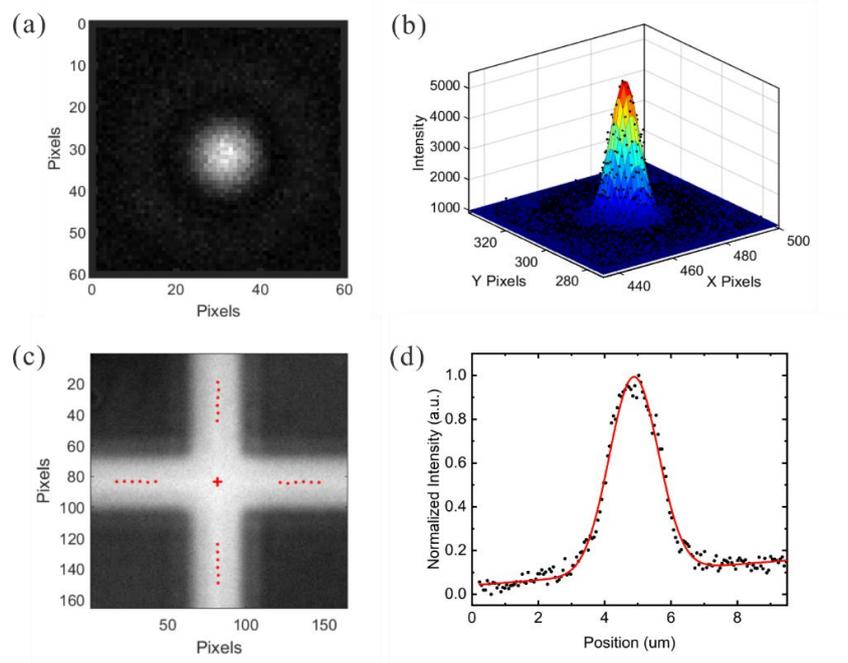

Supplementary Fig. 3

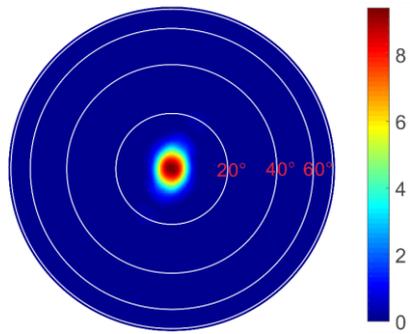

Supplementary Fig. 4

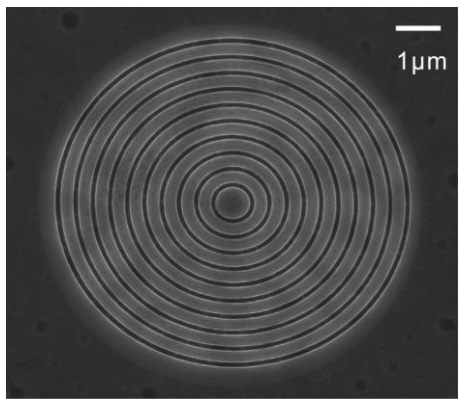

Supplementary Fig. 5

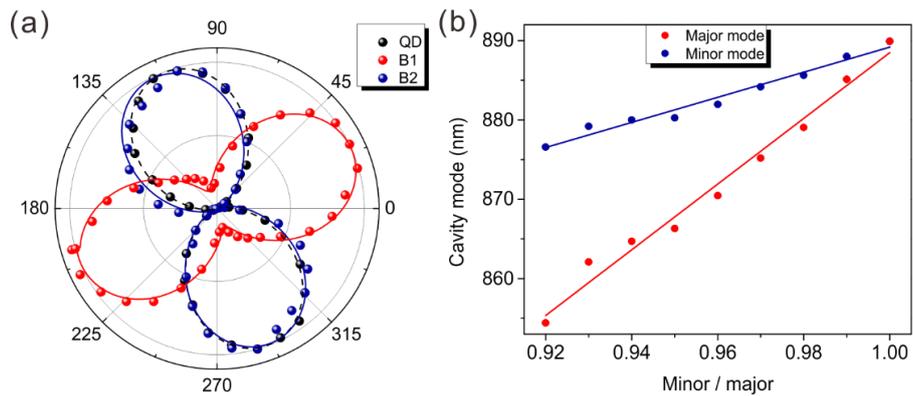

Supplementary Fig. 6

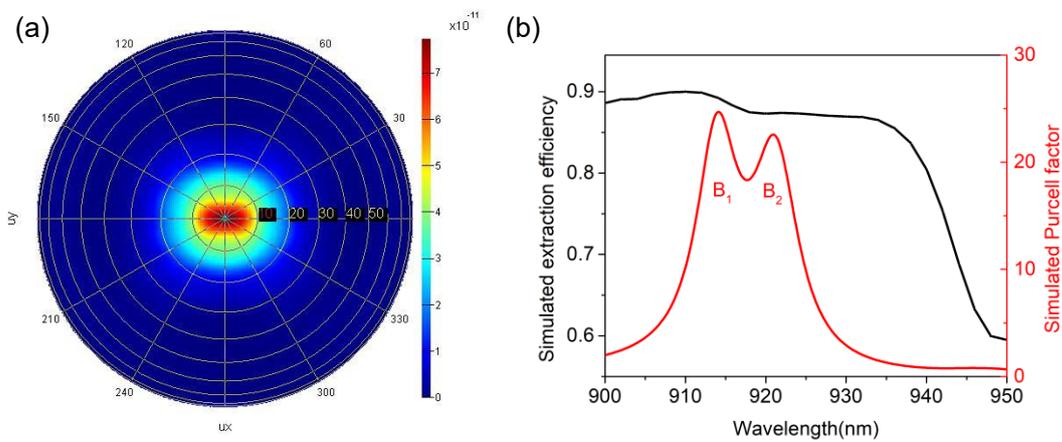

Supplementary Fig. 7

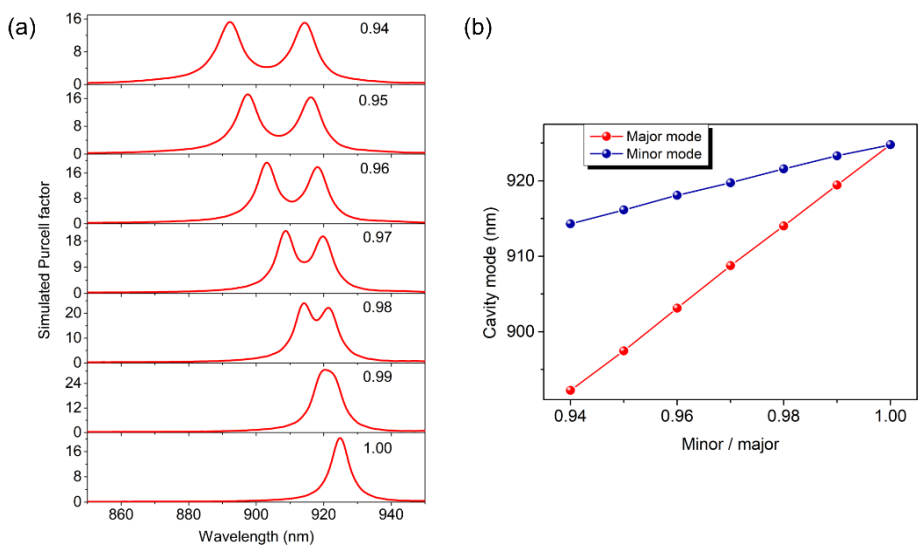

Supplementary Fig. 8

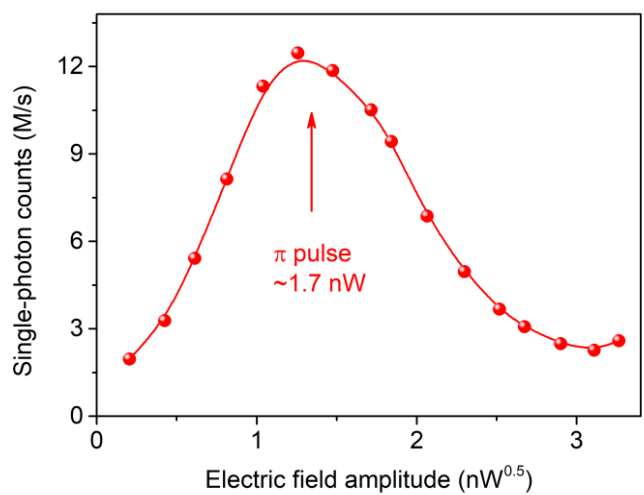

Supplementary Fig. 9

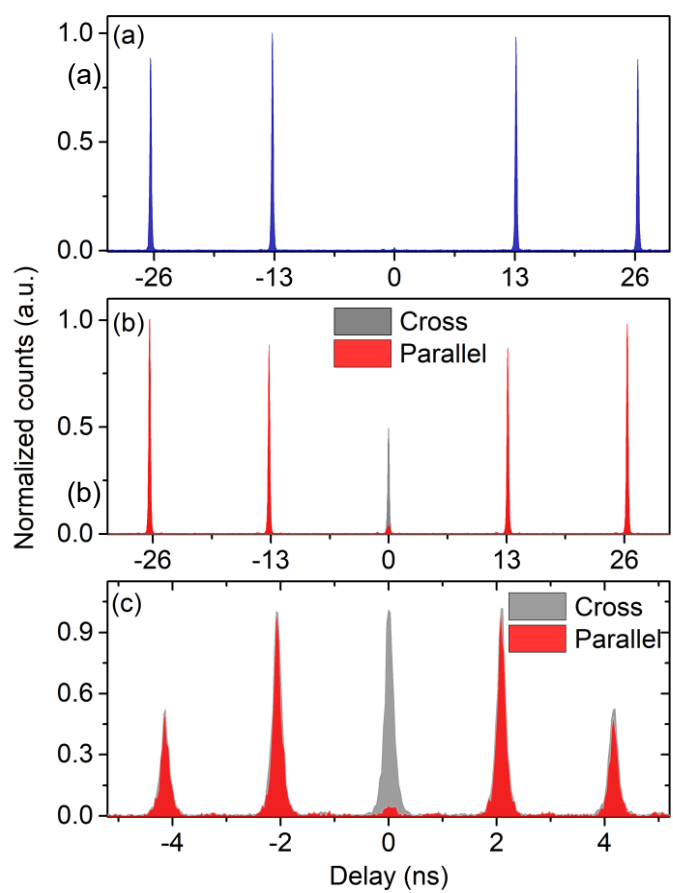

Supplementary Fig. 10

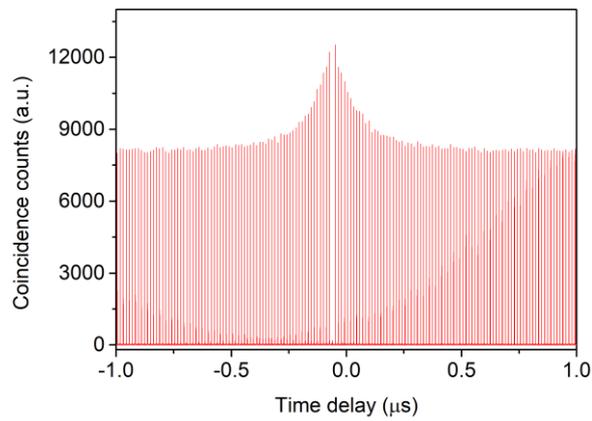

Supplementary Fig. 11

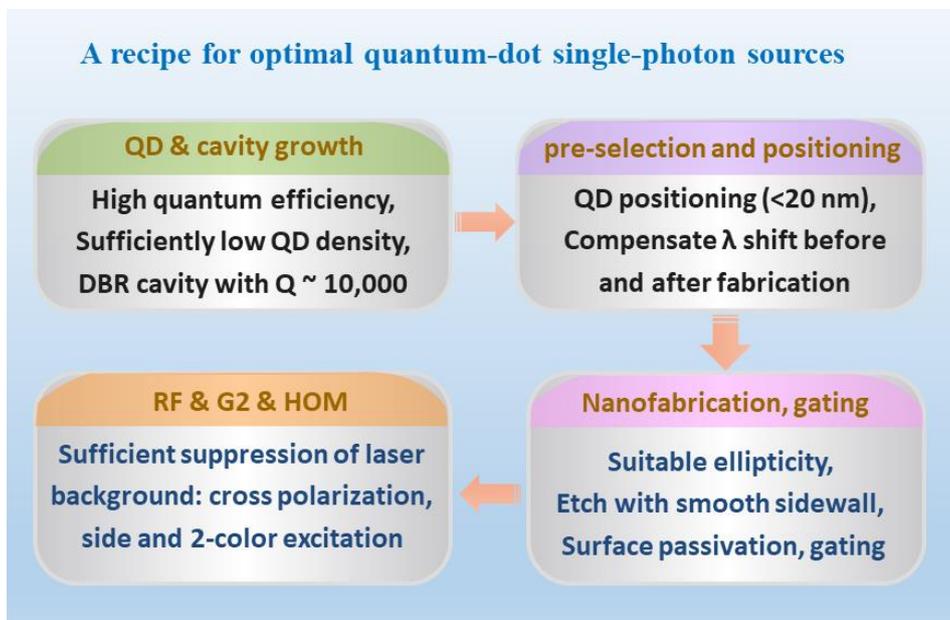

Supplementary Fig. 12